\documentclass[aps, prl, twocolumn, english, notitlepage, showkeys
]{revtex4-2}

\usepackage{graphicx}
\usepackage{dcolumn}
\usepackage{bm}
\usepackage[colorlinks=true, linkcolor=blue, citecolor=blue, urlcolor=blue]{hyperref}

\usepackage{mhchem}
\usepackage{amsmath}
\usepackage{physics}
\usepackage{makecell}
\begin{document}

\preprint{APS/123-QED}

\title{Double-Bridge Mechanism for Enhancing $T_c$ in Oxide Superconductors}

\author{Jun-jie Shi\textsuperscript{1,$^*$}, Juan Du\textsuperscript{2}, and Yao-hui Zhu\textsuperscript{3}}

\affiliation{\textsuperscript{1}State Key Laboratory for Artificial Microstructures and Mesoscopic Physics, School of Physics, Peking University Yangtze Delta Institute of Optoelectronics, Peking University, Beijing 100871, China}
\affiliation{\textsuperscript{2}School of Physics and Optoelectronic Engineering, Beijing University of Technology, Beijing 100124, China}
\affiliation{\textsuperscript{3}Physics Department, Beijing Technology and Business University, Beijing 100048, China}

\date{\today}

\begin{abstract}
We propose a new double-bridge mechanism to significantly enhance $T_c$ in ionic oxide superconductors. Based on our recently proposed ionic-bond-driven O/Cu-bridged (bridge-I) pairing \textbf{e$^-$-O-e$^-$}/\textbf{h$^+$-Cu-h$^+$} formed in the pseudogap phase ($T_c<T<T^*$), we reveal a key bridge-II Cu/O-mediated inter-pair attraction that overcomes direct Coulomb repulsion and drives coherent Bose-Einstein condensation (BEC) of preformed Cooper pairs. Within the BEC framework (Eq.~(\ref{eqn:3})), $T_c$ follows the Uemura scaling $(n_{\rm pair}^{\rm 3D})^{2/3}/m_{\rm pair}^*$ or $n_{\rm pair}^{\rm 2D}/m_{\rm pair}^*$ and increases linearly with the attractive scattering length $a<0$. Strengthening bridge-II attraction, minimizing $m_{\rm pair}^*$, and optimizing $n_{\rm pair}^{\rm 3D}$ are the key to maximizing $T_c$. This double-bridge mechanism unifies the \textbf{eV-scale} strong pairing at room temperature and BEC, provides a universal route toward higher $T_c$, and guides the design of next-generation superconductors.
\end{abstract}

\maketitle
Since the discovery of high-$T_c$ superconductivity in cuprates in 1986~\cite{Bednorz1986}, the field has remained one of the most compelling and  challenging in condensed-matter physics. Realizing superconductivity at higher temperatures, including room temperature, remains a central goal. Despite extensive work, a universally accepted microscopic mechanism remains elusive, and research has branched into numerous systems, including cuprates~\cite{Mourachkine2002}, nickelates~\cite{Wang2024,WangNingning2024}, iron-based superconductors~\cite{Hideo2018}, multi-hydrogen superconductors under ultrahigh-pressure ($\sim$200 GPa)~\cite{Gao2021}, organic superconductors~\cite{ishiguro1998}, heavy-fermion superconductors~\cite{RevModPhys.1984}, kagome superconductors~\cite{Ortiz2021,Xu2025}, `magic angle' twisted bilayer graphene (MATBG)~\cite{Cao2018}, and so on. To date, no clear consensus exists on the most promising materials platform or the fundamental principles for reliably raising $T_c$.

For practical applications, superconductors require high $T_c$, high critical magnetic field, high critical current, good ductility, and ambient-pressure stability and superconductivity---the so-called ``3 highs + ductility + ambient pressure'' criteria. Unfortunately, no known material satisfies all these conditions simultaneously. Progress urgently demands a unified theoretical framework to guide material design. As emphasized by Norman~\cite{Norman2016}, any research disconnected from practical goals may have no future.

Against this background and by considering the strongest interaction and dominance of \textbf{eV-scale} ionic bonding, affinity of O$^-$ (1.46 eV) and O$^{2-}$ (-8.08 eV) and large two-electron ionization energy ($\sim$15-28 eV) of metal atoms in oxide superconductors~\cite{Atkins2010,Kittel2005}, we recently proposed a transformative idea of electron e$^-$ (hole h$^+$) pairing bridged by oxygen O (metal M) atoms named as bridge-I, i.e., the ionic-bond-driven \textbf{e$^-$-O-e$^-$} (\textbf{h$^+$-M-h$^+$}) itinerant Cooper pairing formed in the pseudogap phase ($T_c<T<T^*$, $T^*\geq 300$ K~\cite{Mourachkine2002,Uemura2003,Basov2005})~\cite{shi2025}. This picture is built on the principle of ``tracing electron footprints to explore pairing mechanisms'' and the solid foundation of the chemical-bond$\rightarrow$structure$\rightarrow$property relationship. It applies universally to all ionic-bonded superconductors including cuprates, nickelates, and iron-based superconductors.

Here, we use this universal pairing image to explore new methods for enhancing $T_c$ by combining the layered crystal structure of high-$T_c$ superconductors~\cite{Park1995} and the 4$a$-periodic Cooper-pair charge stripe phase~\cite{Kohsaka2007,Vojta2008}. Through detailed analysis of the dominant interactions between preformed Cooper pairs, we establish a new microscopic route to enhance $T_c$ within the Bose-Einstein condensation (BEC) framework (Eq.~(\ref{eqn:3}))~\cite{Pitaevskii2016,Chen-RevModPhys-2024}, i.e., increasing the attraction between Cooper pairs and reducing their effective mass under the optimal carrier concentration. These results establish a unified, chemically intuitive, and material-predictive framework for high-$T_c$ superconductivity in ionic oxides.

\begin{figure}
    \centering  \includegraphics[width=\linewidth]{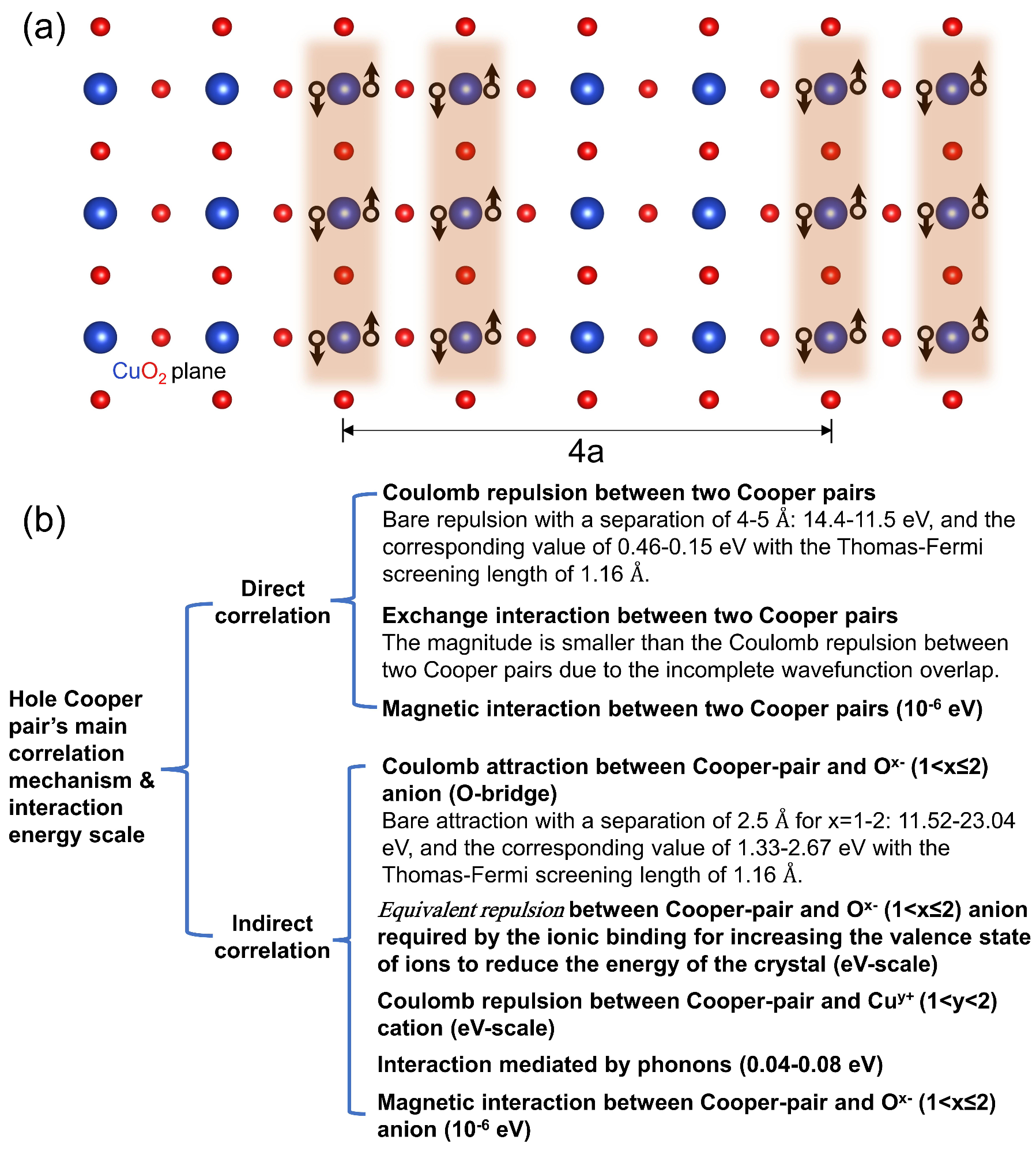}
    \caption{(a) The 4$a$-period Cu-bridged (\textbf{bridge-I}) hole Cooper-pair \textbf{\ce{h^+}-Cu-\ce{h^+}} stripe phase in the \ce{CuO2}  plane~\cite{Kohsaka2007,Vojta2008,shi2025}. Neighboring Cooper pairs are mutually attracted via an O$^{x-}$ ($1<x\leq 2$) anion acting as \textbf{bridge-II}. (b) Dominant direct and O-bridged correlation mechanisms between two \textbf{\ce{h^+}-Cu-\ce{h^+}} hole Cooper pairs and their corresponding energy scales. Analogous correlation mechanisms apply to the O-bridged \textbf{\ce{e^-}-O-\ce{e^-}} electron Cooper pairs, with Cu cations serving as bridge-II.}
    \label{fig:1}
\end{figure}

\begin{figure*}
    \centering  \includegraphics[width=\linewidth]{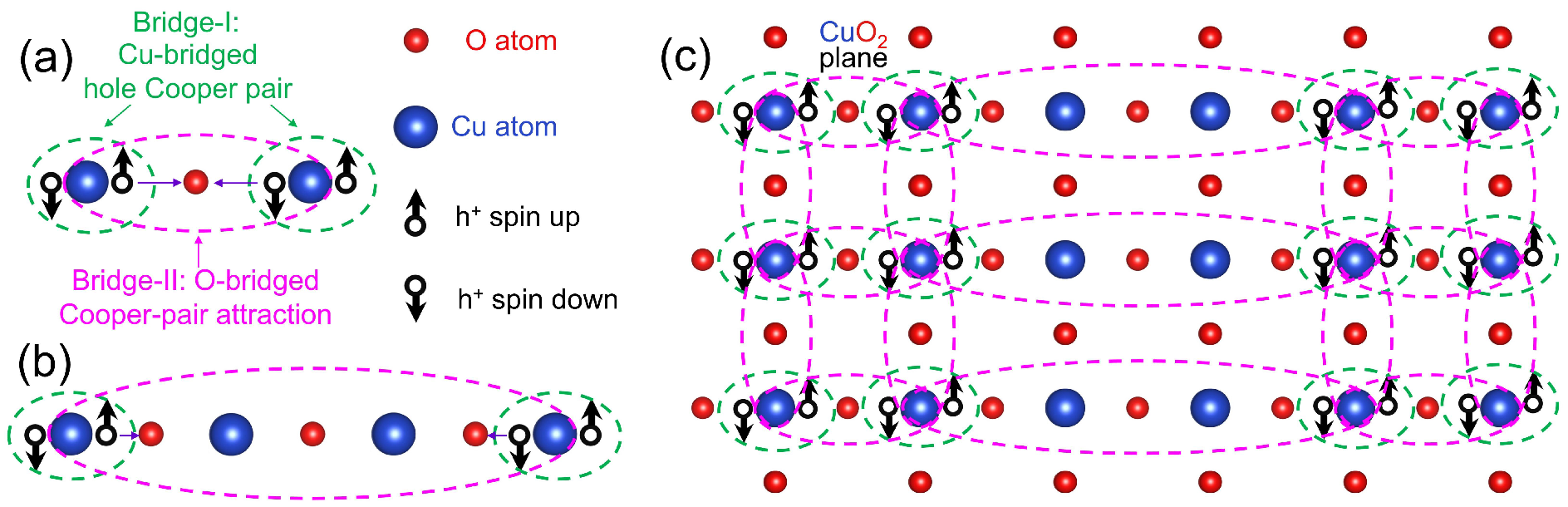}
    \caption{The \textbf{double-bridge mechanism} for high-$T_c$ superconductivity in cuprates. (a) Strong Coulomb attraction from the O$^{x-}$ ($1<x\leq 2$) \textbf{bridge-II} anion acting on two adjacent Cu-bridged (\textbf{bridge-I}) \textbf{\ce{h^+}-Cu-\ce{h^+}} hole Cooper pairs, combined with strong Coulomb repulsion between Cu$^{y+}$ ($1<y<2$) cations and nearby Cooper pairs, pushes \textbf{\ce{h^+}-Cu-\ce{h^+}} pairs toward neighboring O$^{x-}$ anions, yielding an effective eV-scale strong indirect mutual attraction between the two \textbf{\ce{h^+}-Cu-\ce{h^+}} pairs. In the pseudogap phase with $T_c<T<T^*$ ($T^*$: the pairing-onset temperature), this indirect attraction balances the direct Coulomb repulsion and the effective repulsion between O$^{x-}$ anions and \textbf{\ce{h^+}-Cu-\ce{h^+}} pairs imposed by ionic bonding to lower the crystal energy and maintain structural stability (Fig.~\ref{fig:1}(a)~\cite{shi2025}, Fig.~\ref{fig:1}(b)). For $T\leq T_c$, the reduced kinetic energy of electrons surrounding O$^{x-}$ weakens their ability to escape oxygen nuclei, slightly increasing the negative charge $Q$ on O$^{x-}$ as $Q\rightarrow Q+\delta Q$ ($\delta Q\ll Q$), which is further enhanced by the transition of Cooper pairs from excited to ground states (see text). Likewise, the positive valence of Cu$^{y+}$ cations increases modestly below $T_c$, breaking the balance. The Coulomb attraction between O$^{x-}$ and neighboring \textbf{\ce{h^+}-Cu-\ce{h^+}} pairs is thereby strengthened, producing a net inter-pair attraction that drives the BEC of \textbf{\ce{h^+}-Cu-\ce{h^+}} pairs and enhances $T_c$ (Eq.~(\ref{eqn:3}))~\cite{Pitaevskii2016,Chen-RevModPhys-2024}. (b) For two \textbf{\ce{h^+}-Cu-\ce{h^+}} pairs separated by several lattice constants along the Cu--O--Cu chain, their strong local attractions to nearby O$^{x-}$ anions dominate over distant interactions, creating an effective indirect mutual attraction. At $T\leq T_c$, these separated pairs also undergo BEC via O-bridge-II coupling. (c) Combining panels (a) and (b) and accounting for the hole stripe phase, all bridge-I \textbf{\ce{h^+}-Cu-\ce{h^+}} Cooper pairs in the CuO$_2$ planes ``hold hands'' via bridge-II oxygen anions and condense coherently into the superconducting state---the double-bridge mechanism of high-$T_c$ superconductivity. This unified and beautiful picture reveals the microscopic mechanism of high-$T_c$ superconductivity. For O-bridged (bridge-I) \textbf{\ce{e^-}-O-\ce{e^-}} electron Cooper pairs, an analogous double-bridge scheme holds with Cu$^{y+}$($1<y<2$) cations acting as bridge-II to drive their BEC. A similar double-bridge mechanism applies to high-$T_c$ nickelates via the NiO$_2$ plane~\cite{Puphal2025}.}
    \label{fig:2}
\end{figure*}

\textit{Correlation of Cooper pairs and their interaction energy scale}---Before delving into the BEC of Cooper pairs in high-$T_c$ oxides~\cite{Pitaevskii2016,Chen-RevModPhys-2024}, we first analyze the dominant interaction mechanisms between two Cooper pairs. Figure~\ref{fig:1}(a) shows that two Cu-bridged (\textbf{bridge-I}) \textbf{\ce{h+}-Cu-\ce{h+}} hole Cooper pairs exhibit not only direct interactions but also indirect interactions mediated by oxygen anions (\textbf{bridge-II}). This combined picture constitutes the \textbf{double-bridge mechanism} of high-$T_c$ superconductivity (see Fig.~\ref{fig:2} for details). A localized lattice-constant-scale \textbf{\ce{h+}-Cu-\ce{h+}} hole Cooper pair carries a charge of $+2e$, has $d$-wave symmetry with the angular momentum quantum number $l$=2, and is a spin-zero boson ($s$=0)~\cite{shi2025}. Direct Coulomb repulsion exists between two \textbf{\ce{h+}-Cu-\ce{h+}} pairs, with a minimum separation set by the lattice constant $a\approx 4$--$5$ {\AA} of the CuO$_2$ plane~\cite{Mourachkine2002}, corresponding to a bare repulsion energy of 14.4-11.5 eV. Under strong screening with an electron concentration of $1\times 10^{21}$ cm$^{-3}$, the Thomas--Fermi screening length is about 1.16 {\AA}, and the screened Coulomb repulsion drops exponentially to 0.46-0.15 eV (Fig.~\ref{fig:1}(b)). Exchange interactions also arise from wavefunction overlap, but their magnitude is noticeably weaker than the direct Coulomb repulsion due to incomplete overlap~\cite{Sakurai1994}. Treating the $d$-wave \textbf{\ce{h+}-Cu-\ce{h+}} pair as a magnetic dipole with orbital moment $M_z=2\mu_{\rm B}$, the magnetic interaction is estimated to be on the order of 10$^{-6}$ eV, which is negligible compared with the Coulomb interaction.

In addition to direct Coulomb repulsion, oxygen anions act as correlation bridges that link neighboring \textbf{\ce{h+}-Cu-\ce{h+}} Cooper pairs (Fig.~\ref{fig:1}(a)), and \textbf{eV-scale} strong Coulomb attraction exists between each  anion and its nearest Cooper pairs. As shown previously~\cite{shi2025}, the oxygen anion has an average valence state $1<x\leq 2$ and lies roughly half a lattice constant away from the Cooper pairs, i.e., $a/2\approx 2.5$ {\AA}. The bare Coulomb attraction between a Cooper pair ($+2e$) and O$^{x-}$ ($-e$ or $-2e$) is thus estimated to range from 11.52 to 23.04 eV. Under strong screening with a Thomas--Fermi screening length of 1.16 {\AA}~\cite{shi2025}, this attraction falls exponentially to 1.33-2.67 eV (Fig.~\ref{fig:1}(b)), which remains roughly one order of magnitude larger than the direct inter-pair Coulomb repulsion.

Furthermore,  an \textit{equivalent \textbf{eV-scale} repulsion} exists between each \textbf{\ce{h+}-Cu-\ce{h+}} pair and O$^{x-}$, which arises from ionic bonding constraints that raise ion valences to lower the total crystal energy and preserve structural stability (Fig. 1(a)~\cite{shi2025}). A similar\textbf{ eV-scale} Coulomb repulsion also acts between \textbf{\ce{h+}-Cu-\ce{h+}} pairs and Cu$^{y+}$ ($1<y<2$)  cations. For comparison, the electron-phonon coupling strength in high-$T_c$ cuprates is experimentally found to be only 0.04-0.08 eV~\cite{Yan2023,Lanzara2001}. Assuming a magnetic moment of 0.2$\mu_{\rm B}$ for O$^{x-}$~\cite{Rajan2023}, the magnetic interaction energy between O$^{x-}$ and \textbf{\ce{h+}-Cu-\ce{h+}} is on the order of 10$^{-6}$ eV, which is negligible. An analogous analysis applies to O-bridged (bridge-I) \textbf{\ce{e-}-O-\ce{e-}} electron Cooper pairs.

\textit{Double-bridge mechanism of high-$T_c$ superconductivity}---Comprehensively considering all dominant \textbf{eV-scale} correlation mechanisms summarized in Fig.~\ref{fig:1}(b), we establish an elegant \textbf{double-bridge mechanism} for high-$T_c$ superconductivity in cuprates, as illustrated in Fig.~\ref{fig:2}. The Cu-bridge (\textbf{bridge-I}) forms \textbf{\ce{h+}-Cu-\ce{h+}} Cooper pairs~\cite{shi2025}. The O-bridge (\textbf{bridge-II}) provides an effective inter-pair attraction that overcomes direct Coulomb repulsion and drives coherent BEC of Cooper pairs across the entire CuO$_2$ plane~\cite{Pitaevskii2016,Chen-RevModPhys-2024}, directly increasing the BEC critical temperature $T_{\rm BEC}$, i.e., the superconducting transition temperature $T_c$, according to the interacting BEC formula Eq.~(\ref{eqn:3}). Grounded in the fundamental nature of ionic bonding in high-$T_c$ superconductors, this mechanism is universal and can be extended to strongly ionic superconductors including cuprates, nickelates, iron-based systems, and other promising ionic compounds, offering a concrete guideline for further enhancing $T_c$. An analogous BEC scenario can be constructed for O-bridged \textbf{\ce{e-}-O-\ce{e-}} electron Cooper pairs.

\begin{figure}
    \centering  \includegraphics[width=6.0 cm]{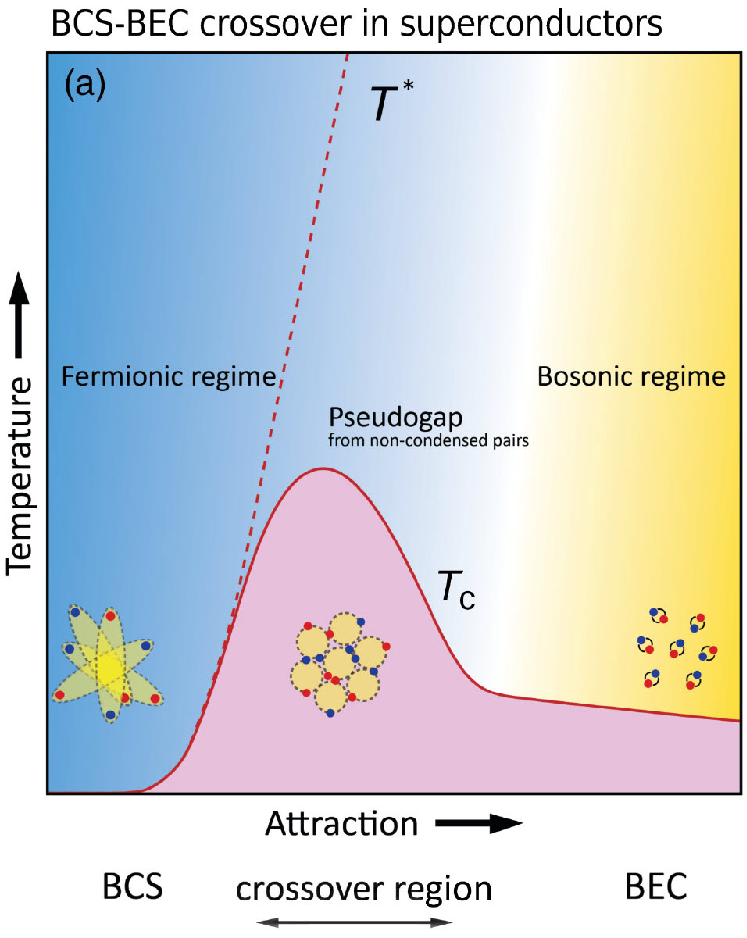}
    \caption{BCS--BEC crossover phase diagram governed by intra-pair attraction in superconductors~\cite{Chen-RevModPhys-2024}.}
    \label{fig:3}
\end{figure}

\textit{Cooper pair transition to ground state: enhanced attraction with bridge-II atoms and superconducting fluctuations}---To elucidate how Cooper-pair transitions affect the attractive interaction between bridge-II atoms and Cooper pairs, we focus on hole carriers, the dominant charge carriers in high-$T_c$ cuprates. According to our previously proposed \textbf{eV-scale} ionic-bond-driven atom-bridged (bridge-I) room-temperature hole-pairing mechanism~\cite{shi2025}, strong ionic bonds in the crystal overcome Coulomb repulsion between holes and form excited-state \textbf{\ce{h+}-M-\ce{h+}} hole Cooper pairs in the pseudogap phase ($T_c<T<T^*$). As the temperature is lowered to $T_c$, these Cooper pairs release the superconducting gap energy $E_{sc}$, undergo BEC (see Fig.~\ref{fig:3}), and transition from excited to ground states, entering superconducting phase. For high-$T_c$ Bi$_2$Sr$_2$CaCu$_2$O$_{8+\delta}$ (Bi2212), $E_{sc}\approx 40$ meV~\cite{Hufner2008}. Where does this released energy go? What impact will it have on superconductivity?

Based on the atomic bonding scenario in high-$T_c$ oxides (Fig.~1~\cite{shi2025}), the total energy released by Cooper pair transition from excited to ground states promotes weak ionization of a small number of metal cations. As required by the crystal binding, the ionized electrons then accumulate toward neighboring oxygen anions driven by \textbf{eV-scale} strong ionic bonds, slightly increasing the negative valence of O$^{x-}$ to O$^{(x+\delta x)-}$ (Fig.~\ref{fig:2}), which induces local superconducting fluctuations around partial oxygen anions. The detailed microscopic process requires further investigation. Obviously, the enhanced valence of O$^{(x+\delta x)-}$ strengthens their Coulomb attraction with adjacent \textbf{\ce{h+}-Cu-\ce{h+}} hole Cooper pairs, leading to a net mutual attraction between Cooper pairs and further increasing $T_c$ (see Eq.~(\ref{eqn:3})). As a consequence of local superconducting fluctuations around a small number of oxygen anions, the energy released during pair condensation contributes to an enhanced $T_{c}^{\rm onset}$ in oxide superconductors~\cite{WangNingning2024}. An analogous picture holds for O-bridged \textbf{\ce{e-}-O-\ce{e-}} electron Cooper pairs.

\textit{Intra-pair attraction: BCS-to-BEC transition and BEC in oxides}---Quite recently, spectroscopic-imaging scanning tunneling microscopy has been used to probe the electronic structure of \ce{CuO2} planes in an electron-doped infinite-layer cuprate Sr$_{1-x}$Nd$_x$CuO$_2$.~\cite{Zhu2025}. The results strongly indicate the appearance of BEC-like behavior preceding macroscopic superconducting condensation, and provide direct real-space evidence for BCS--BEC crossover in \ce{CuO2} planes due to a negative chemical potential $\mu<0$~\cite{Zhu2025,Melo1993}. Given that the $T_c$ dome with overlain coherence length provides an ideal prototype for BCS--BEC crossover physics, Ref.~\cite{Chen-npj-2024} confirmed the presence of such crossover in high-$T_c$ cuprates. The BEC of Cooper pairs in high-$T_c$ cuprates has also been supported by the Uemura plots~\cite{Uemura1989,Uemura1991,Uemura2003,Uemura2004} and previous works~\cite{Mourachkine2002,Cheng2016,Bozovic2017}.

We know from BCS--BEC crossover phase diagram (Fig.~\ref{fig:3}) that weak intra-pair attraction leads to large overlapping Cooper pairs, $T^*\approx T_c$, and the absence of a pseudogap in the BCS regime. Moderate intra-pair attraction yields smaller, tighter pairs, $T^*>T_c$, and the emergence of the pseudogap phase in the crossover regime. Strong intra-pair attraction results in compact, non-overlapping bosonic pairs with $T^*\gg T_c$ in the BEC regime. As intra-pair attraction increases, Cooper pairs shrink and acquire more bosonic character, and the system evolves continuously from BCS to BEC. Unlike conventional BCS superconductors, which possess large Cooper-pair size of 400--$10^4$ {\AA} due to weak electron-phonon coupling~\cite{Mourachkine2002}, Cooper pairs in oxide superconductors, driven by \textbf{eV-scale} strong ionic bonds, are localized around bridge-I atoms with a small size comparable to the lattice constant, just as independent, stable bosonic molecules~\cite{shi2025}. Furthermore, a large pseudogap (e.g., $\sim$80 meV in Bi2212~\cite{Hufner2008}) has been widely observed in oxide superconductors. These results clearly indicate that ionic-bonded oxide superconductors exhibit BEC superconductivity rather than BCS superconductivity, confirmed by the aforementioned experiments~\cite{Zhu2025,Melo1993,Chen-npj-2024,Uemura1989,Uemura1991,Uemura2003,Uemura2004,Mourachkine2002,Cheng2016,Bozovic2017}.

\begin{figure}
    \centering  \includegraphics[width=\linewidth]{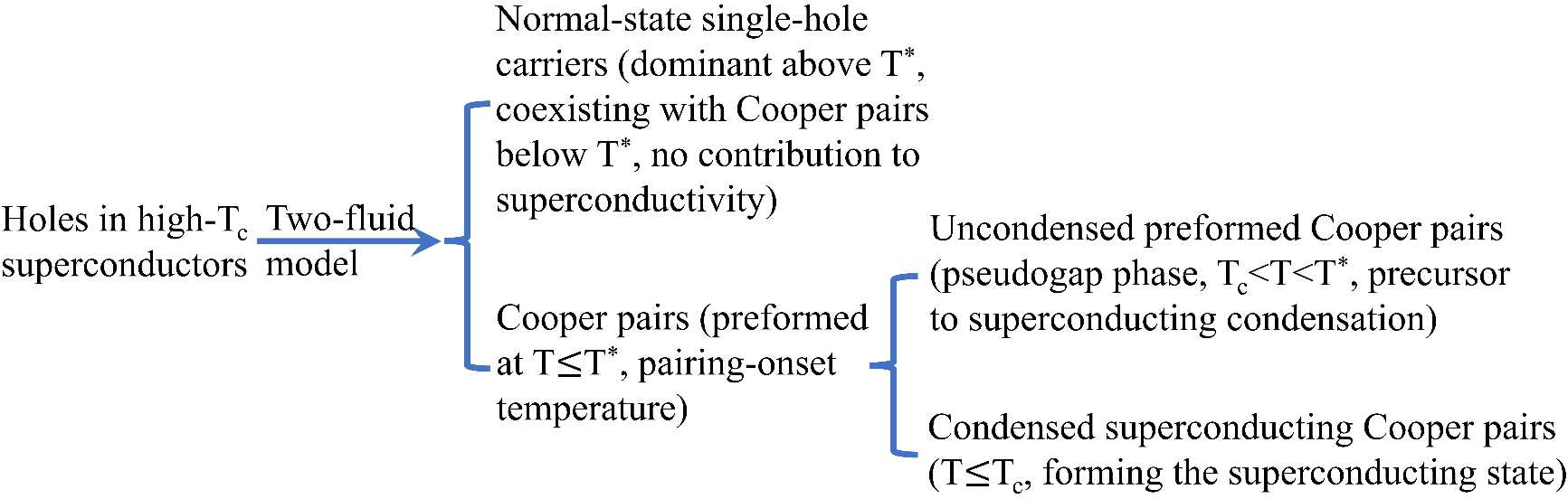}
    \caption{Classification of conductive charge carriers in high-$T_c$ superconductors: single-hole carriers in the normal state, preformed Cooper pairs in the pseudogap phase, and condensed Cooper pairs responsible for superconductivity. This classification applies equally to electron pairs.}
    \label{fig:4}
\end{figure}

\textit{BEC of interacting Cooper pairs and the $T_c$ formula}---It is well established that high-$T_c$ cuprates exhibit characteristic BEC-like behaviors~\cite{Mourachkine2002,Uemura2003,Chen-RevModPhys-2024,Zhu2025,Melo1993,Chen-npj-2024,Bozovic2017}, including \textbf{eV-scale} strong-attraction Cooper pairing, small pair sizes, strong pairing fluctuations, preformed pairs in the pseudogap phase ($T_c<T<T^*$), two energy gaps (pseudogap and superconducting gap), the dome-shaped superconducting phase diagram, strong charge localization, and dominant ionic bonding~\cite{Mourachkine2002,shi2025,Hufner2008}. These preformed Cooper pairs are spin-zero bosons that can undergo a BEC phase transition from a Bose liquid into the superconducting state (Fig.~\ref{fig:3})~\cite{Chen-RevModPhys-2024,Mourachkine2002,shi2025,Cheng2016,Uemura1989,Uemura1991,Uemura2003,Uemura2004,Bozovic2017}. The concept of BEC was originally introduced by Einstein in 1925, and its foundational theory was well developed prior to 1995~\cite{Pitaevskii2016}. Within the well-known two-fluid model~\cite{Sharma2021}, the conductive charge carriers $N$ in the superconducting phase ($T\leq T_c$) are partitioned into condensed superconducting Cooper pairs N$_{\rm s}$, uncondensed pseudogap-phase Cooper pairs N$_{\rm p}$, and uncondensed normal-state single carriers N$_{\rm n}$ (see Fig.~\ref{fig:4}). We thus have the following total itinerant carrier density,
\begin{equation}
\label{eqn:1}
\begin{aligned}
 n=\frac{N}{V}=\frac{N_{\rm s}+N_{\rm p}+N_{\rm n}}{V}=n_{\rm s}+n_{\rm p}+n_{\rm n},
    \end{aligned}
\end{equation}
where $V$ is volume, $n_{\rm s}$=$N_{\rm s}/V$ is the superconducting (single-particle) carrier density, $n_{\rm p}$=$N_{\rm p}/V$ is the uncondensed carrier density in the pseudogap phase, and $n_{\rm n}$=$N_{\rm n}/V$ is the density of normal-state single carriers. For noninteracting ideal Bose Cooper pairs, the BEC critical temperature $T_{\rm BEC}$ i.e., the superconducting transition temperature $T_c^0$, is given by~\cite{Pitaevskii2016},
\begin{equation}
\label{eqn:2}
\begin{aligned}
 T_c^0 =\frac{2 \pi \hbar^2}{m_{\rm pair}^* k_{\rm B}}\left(\frac{n_{\rm pair}^{\rm 3D}}{2.612}\right)^{2/3}\propto\frac{\left(n_{\rm pair}^{\rm 3D}\right)^{2/3}}{m_{\rm pair}^*},
 \end{aligned}
\end{equation}
where $k_{\rm B}$ is Boltzmann's constant, $\hbar$ is the reduced Planck's constant, $m_{\rm pair}^*$ is the effective mass of a Cooper pair, and $n_{\rm pair}^{\rm 3D}$=$n_{\rm s}/2$ is the total density of condensed Cooper pairs responsible for superconductivity. The superscript ``0'' denotes the ideal noninteracting Cooper pairs. Equation~(\ref{eqn:2}) is supported by the Uemura plot---a universal linear scaling of $T_c$ with $n_{\rm pair}^{\rm 2D}/m_{\rm pair}^*$ or $(n_{\rm pair}^{\rm 3D})^{2/3}/m_{\rm pair}^*$ for underdoped quasi-two-dimensional (Q2D) layered and three-dimensional (3D) unconventional superconductors~\cite{Uemura1989,Uemura1991,Uemura2003,Uemura2004}. It shows that $T_c^0$ is inversely proportional to $m_{\rm pair}^*$ and directly proportional to $(n_{\rm pair}^{\rm 3D})^{2/3}$ or $n_{\rm pair}^{\rm 2D}$, in agreement with experiments~\cite{Cao2018,Bozovic2017}. This explains why heavy-fermion superconductors, with electron effective mass $m^*\sim 200m_e$, exhibit very low $T_c<1$ K~\cite{RevModPhys.1984,Dong2025}. In contrast to high-$T_c$ cuprates with total carrier density $n\sim 10^{21}$ cm$^{-3}$~\cite{Mourachkine2002,Harshman1992}, the much lower density $\sim 10^{17}$--$10^{18}$ cm$^{-3}$ in MATBG leads to its low $T_c\sim 1.7$ K~\cite{Cao2018,Chen-RevModPhys-2024}. The high $T_c$ in cuprates arises from room-temperature preformed pairing, guaranteed by the \textbf{eV-scale} large pairing energy and stable crystal structure at room temperature, and electronic BEC with $m_{\rm pair}^*\sim 10m_e$ and $n_{\rm pair}^{\rm 3D}\sim 10^{19}$--$10^{20}$ cm$^{-3}$ (Table~\ref{tbl:1})~\cite{shi2025,Bozovic2017,Harshman1992}. To maximize $T_c$, $m_{\rm pair}^*$ should be minimized while maintaining the optimal Cooper-pair density $n_{\rm pair}^{\rm 3D}$, as dictated by the dome-shaped superconducting phase diagram in high-$T_c$ cuprates~\cite{Mourachkine2002}.

The superconducting transition temperature $T_c$, corresponding to the BEC critical temperature for interacting Cooper pairs, can be obtained within the mean-field method as follows~\cite{Pitaevskii2016},
\begin{equation}
\label{eqn:3}
\begin{aligned}
    T_c=T_c^0 \left(1-3.426\frac{a}{\lambda_0}\right),
    \end{aligned}
\end{equation}
where $a$ is the scattering length characterizing the inter-pair interaction strength, with $a<0$ for attractive interactions and $a>0$ for repulsive interactions, respectively. Typically, $|a|$ can exceed the inter-pair distance. Equation~(\ref{eqn:3}) clearly shows that $T_c$ increases linearly with $|a|$ for attractive inter-pair interactions, paving a promising route to significantly enhance $T_c$ for BEC of preformed Cooper pairs toward higher-temperature superconductivity. Such inter-pair attraction is thus highly beneficial for maximizing $T_c$. Therefore, \textbf{a complete and consistent high-$T_c$ theory must address two key issues: (1) a strong attractive pairing mechanism between holes (electrons), as demonstrated in Ref.~\cite{shi2025}; (2) an inter-pair attraction mechanism that drives coherent condensation of preformed Cooper pairs and enhances $T_c$, as confirmed in Fig.~\ref{fig:2}.} As a remaining open challenge, maximizing $|a|$ represents a critical task in materials design.

The thermal de Broglie wavelength $\lambda_0$ at the critical temperature $T_c^0$, which depends only on the Cooper-pair density
$n_{\rm pair}^{\rm 3D}$, is given by:
\begin{equation}
\label{eqn:4}
\begin{aligned}
    \lambda_0=\frac{\sqrt{2\pi} \hbar}{\sqrt{m_{\rm pair}^* k_{\rm B}T_c^0}}=
& \left(\frac{2.612}{n_{\rm pair}^{\rm 3D}}\right)^{1/3}.
    \end{aligned}
\end{equation}
Equations~(\ref{eqn:2})-(\ref{eqn:4}) show that $T_c$ depends only on the effective mass $m_{\rm pair}^*$, the density $n_{\rm pair}^{\rm 3D}$ of condensed Cooper pairs, and the scattering length $a$. \textbf{$T_c$ can thus be maximized by: (1) optimize doping for optimal carrier density (lock optimal doping); (2) minimizing Cooper pair effective mass (pressure/strain engineering); (3) enhancing inter-pair attraction via bridge-II ions (tune oxygen content; optimize local electrons around in-plane O atoms; tune interlayer ions and charge transfer).} Furthermore, the condensate fraction $f$ of Cooper pairs---defined as the ratio of condensed pairs $N_{\rm s}$/2 to the total number of Cooper pairs  ($N_{\rm s}+N_{\rm p}$)/2---at temperature $T$ is given by~\cite{Pitaevskii2016},
\begin{equation}
\label{eqn:5}
\begin{aligned}
f =\frac{N_{\rm s}}{N_{\rm s}+N_{\rm p}}=
\left\{
\begin{aligned}
& 1-(T/T_c)^{3/2}, & T \leq T_c, \\
& 0, & T > T_c.
\end{aligned}
\right.
    \end{aligned}
\end{equation}

\begin{figure}
    \centering  \includegraphics[width=6 cm]{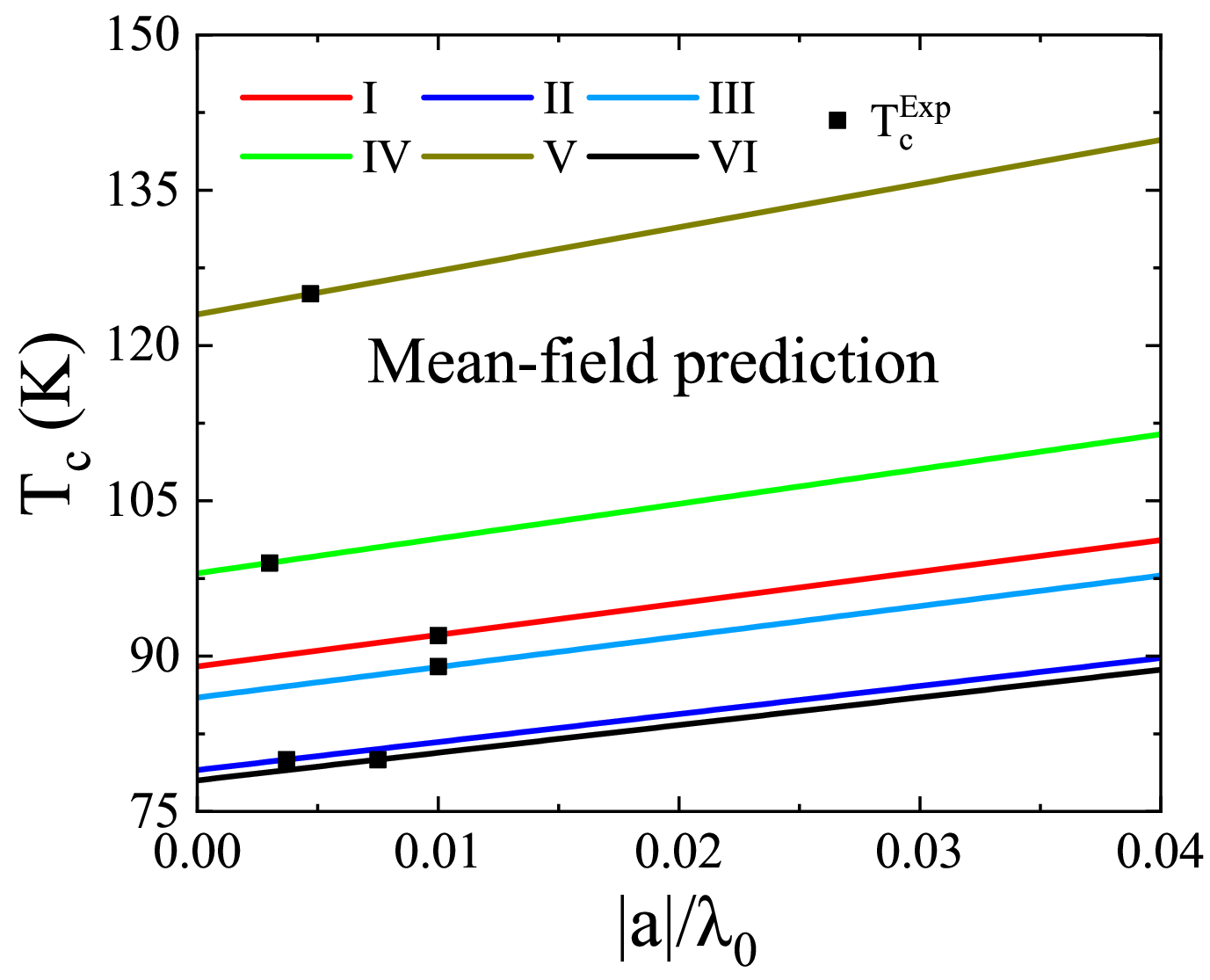}
 \caption{Superconducting critical temperature $T_c$ versus scattering length $|a|$ ($a<0$) for the indirect net attraction between Cu-bridged (bridge-I) \textbf{\ce{h+}-Cu-\ce{h+}} hole Cooper pairs mediated by O-bridge (bridge-II). Results are computed from Eq.~(\ref{eqn:3}) using parameters from Table~\ref{tbl:1} for cuprates I--VI. The figure shows that $T_c$ increases linearly with the attractive inter-pair scattering length $|a|$, demonstrating that strengthening bridge-II--mediated inter-pair attraction is the primary path toward high $T_c$.}
    \label{fig:5}
\end{figure}

Experiments consistently yield a total carrier density $n\sim 10^{21}$ cm$^{-3}$ in high-$T_c$ cuprates~\cite{Mourachkine2002,Wang1987,Harshman1992}.
Since the Cooper-pair density is lower than $n$ (Eq.~(\ref{eqn:1})), a reasonable estimate at finite temperature $T<T_c$ is $n_{\rm pair}^{\rm 3D}\sim 10^{19}$--$10^{20}$ cm$^{-3}$, which is larger than the $10^{18}$ cm$^{-3}$ suggested in Ref.~\cite{Cheng2016}. All relevant material parameters for six representative high-$T_c$ cuprates are summarized in Table~\ref{tbl:1}. The experimental $T_c^{\rm Exp}$ and Cooper-pair effective masses $m_{\rm pair}^*$ are taken from Ref.~\cite{Harshman1992}. The calculated $T_c$ as a function of the scattering length $|a|$ is plotted in Fig.~\ref{fig:5}.

\begin{table}[t]
\centering
\caption{Cooper-pair density $n_{\rm pair}^{\rm 3D}$, effective mass $m_{\rm pair}^*$ (in units of free-electron mass $m_e$), thermal wavelength $\lambda_0$, calculated BEC critical temperature $T_c^0$, and experimental $T_c^{\rm Exp}$ for six cuprates: YBa$_2$Cu$_3$O$_7$ (I), YBa$_2$Cu$_4$O$_8$ (II),  Bi$_2$Sr$_2$CaCu$_2$O$_8$ (III),  Tl$_2$Ba$_2$CaCu$_2$O$_8$ (IV), Tl$_2$Ca$_2$Ba$_2$Cu$_3$O$_{10}$ (V), and  Tl$_{0.5}$Pb$_{0.5}$Sr$_2$CaCu$_2$O$_7$ (VI).}
  \label{tbl:1}
  \begin{ruledtabular}
  \begin{tabular}{lccccc}
    Cuprate & {$n_{\rm pair}^{\rm 3D}$~(10$^{21}$~cm$^{-3}$)} & $m_{\rm pair}^{*a}$ & $\lambda_0$~(nm) & $T_c^0$~(K) & $T_c^{\rm Exp}$~(K)$^a$ \\
\colrule
  I  & 0.622  & 24.0  & 1.613 & 89 & 92 \\
  II  & 0.093  & 7.6  & 3.036 & 79 & 80 \\
  III  & 0.311  & 15.6  & 2.032 & 86 & 89 \\
  IV  & 0.424  & 16.8  & 1.833 & 98 & 99 \\
  V  & 0.331  & 11.4  & 1.991 & 123 & 125 \\
  VI  & 0.071  & 6.4  & 3.330 & 78 & 80 \\
  \end{tabular}
  \end{ruledtabular}
   $^a$From Ref.~\cite{Harshman1992}.
\end{table}

As seen in Fig.~\ref{fig:5}, the BEC temperature $T_c$ for interacting Cooper pairs depends strongly on $n_{\rm pair}^{\rm 3D}$, $m_{\rm pair}^*$, and scattering length $a$. Precise evaluation of these parameters is critical for reliable $T_c$ and remains a major challenge. For ideal noninteracting pairs, $T_c^0\propto (n_{\rm pair}^{\rm 3D})^{2/3}/m_{\rm pair}^*$, as confirmed by Uemura plots~\cite{Uemura1989,Uemura1991,Uemura2003,Uemura2004}. Thus, lowering $m_{\rm pair}^*$ while retaining optimal $n_{\rm pair}^{\rm 3D}$ favors higher $T_c$. In addition, $T_c$ varies linearly with $a$. The bridge-II-mediated attractive interaction between \textbf{\ce{h+}-Cu-\ce{h+}} (\textbf{\ce{e-}-O-\ce{e-}}) Cooper pairs substantially enhances $T_c$, whereas repulsive interactions reduce $T_c$ in a linear fashion.

As previously demonstrated~\cite{shi2025}, Cooper pairs can survive up to room temperature in ionic oxide superconductors, consistent with experimental observations~\cite{Mourachkine2002,Uemura2003,Basov2005} and offering great potential for realizing room-temperature superconductivity. As illustrated in Fig.~\ref{fig:5}, further enhancement of $T_c$ is feasible via rational material parameter optimization and strengthened net inter-pair attraction, despite the inherent challenges. This work establishes a clear and actionable route toward higher-$T_c$ superconductivity: maintaining an optimal Cooper-pair density $n_{\rm pair}^{\rm 3D}$, minimizing the Cooper-pair effective mass $m_{\rm pair}^*$, and enhancing the attractive scattering length $|a|$ ($a<0$).

\textit{Conclusion}---Building on our \textbf{eV-scale} ionic-bond-driven bridge-I pairing scheme for room-temperature \textbf{h$^+$-M-h$^+$} (\textbf{e$^-$-O-e$^-$}) Cooper pairs, we identify the dominant inter-pair interactions in oxide superconductors. We find that bridge-II--mediated indirect Coulomb attraction overwhelms direct repulsion between Cooper pairs and governs their coherent BEC. Based on our double-bridge mechanism (Fig.~\ref{fig:2}), we show that net inter-pair attraction significantly enhances $T_c$ (Eq.~(\ref{eqn:3})). For ionic superconductors, $T_c$ follows the Uemura scaling $(n_{\rm pair}^{\rm 3D})^{2/3}/m_{\rm pair}^*$ or $n_{\rm pair}^{\rm 2D}/m_{\rm pair}^*$ and increases linearly with the attractive scattering length $|a|$ ($a<0$). We identify three key strategies to maximize $T_c$: (1) strengthening bridge-II-mediated inter-pair attraction (tune oxygen content; optimize local electrons around in-plane O atoms; tune interlayer ions and charge transfer), (2) optimizing the Cooper-pair density $n_{\rm pair}^{\rm 3D}$ (lock optimal doping), (3) minimizing the pair effective mass $m_{\rm pair}^*$ (pressure/strain engineering). Our double-bridge mechanism establishes a universal material-oriented pathway and clear design principle to enhance $T_c$  in ionic oxides, with promising prospects toward higher-$T_c$ and even room-temperature superconductivity.

\textit{Acknowledgments}---We thank Prof. Bo Song and H.Q. Luo for fruitful discussions. This work is supported by the National Key Research and Development Program of China (2023YFB4604400).


\textit{Data availability}---The data supporting this study's findings are available within the article.

$^*$Corresponding author:
jjshi@pku.edu.cn

\bibliography{prl-shi}


\end{document}